# Multiple Idiopathic Cervical Root Resorption: A Challenge for a Transdisciplinary Medical-Dental Team

Emily Y. Chu[1], Janina Golob Deeb[2], Brian L. Foster[3], Evlambia Hajishengalis[4], Martha J. Somerman[1], Vivek Thumbigere-Math[1,5]

[1] National Institute of Arthritis and Musculoskeletal and Skin Diseases, National Institutes of Health, Bethesda, MD, USA.
[2] Department of Periodontics, School of Dentistry, Virginia Commonwealth University, Richmond, VA, USA.
[3] Division of Biosciences, College of Dentistry, The Ohio State University, Columbus, OH, USA.
[4] Divisions of Pediatric Dentistry, School of Dental Medicine, University of Pennsylvania, Philadelphia, PA, USA.
5 Division of Periodontology, University of Maryland School of Dentistry, Baltimore, MD, USA.

These authors contributed equally to the manuscript and authors are listed in alphabetical order.

**Corresponding Author:**
Vivek Thumbigere-Math, B.D.S., Ph.D.
Division of Periodontology
University of Maryland School of Dentistry
Baltimore, MD, USA
Email: vthumbigere@umaryland.edu

**Abstract:** 187
**Word count:** 2,989
**Figures:** 1
**Tables:** 1
**References:** 89

**Keywords:** Periodontal Diseases; Periodontium; Osteoclasts; Odontoclasts; Bone Resorption; Genetics; Medications

## ABSTRACT

While tooth root resorption is a normal physiological process required for resorption and exfoliation of primary teeth, root resorption of adult teeth is largely pathological. This perspective focuses on multiple idiopathic cervical root resorption (MICRR), an aggressive form of external root resorption that occurs near the cemento-enamel junction (CEJ). The cause of MICRR remains elusive, however, it is mediated primarily by osteoclasts/odontoclasts. Accumulating case studies and experiments in animal models have provided insights into defining the etiologies and pathophysiological mechanisms for MICRR, which include: systemic conditions and syndromes, inherited genetic variants affecting osteoclast/odontoclast activity, altered periodontal structures, drug-induced root resorption and rebound effects after cessation of anti-resorptive treatment, chemotherapy, exposure to pets or viral infections, and other factors such as inflammatory conditions or trauma. To determine the causative factors, at minimum, a comprehensive health history should be collected for all patients by dental care providers, discussed with other health care providers and appropriate collaborations established. The examples highlighted in this perspective emphasize the need for transdisciplinary research collaborations coupled with integrated management strategies between medicine and dentistry in order to identify cause(s) early and improve clinical outcomes.

# INTRODUCTION

This perspective underscores the need for researchers and clinicians to use transdisciplinary approaches for defining the causes of tooth root resorption, identifying risk factors and developing treatment strategies for pathologically mediated root resorption. In this regard, please see the commentary and short video, 5-7 min, describing the procedure for doing an oral exam for non-dental clinicians published in JAMA, 2018 [1; 2].

While root resorption is a normal physiological process required for resorption and exfoliation of primary teeth [3], root resorption of adult teeth is largely pathological. Pathological root resorption can be broadly classified into internal (i.e., originating within the dental pulp) or external (i.e., attacking the outer root surface) processes, mediated by osteoclasts/odontoclasts [4; 5; 6; 7]. This perspective focuses on multiple idiopathic cervical root resorption (MICRR), an aggressive form of external root resorption that occurs near the cemento-enamel junction (CEJ) (**Fig 1**). The CEJ is the area where the crown transitions into root(s) and where gingival fibers attach to a healthy tooth root and surrounding alveolar bone. As we highlight below, there has been progress in identifying etiologies for MICRR, particularly genetic and medication-associated causes, prompting us to reconsider the term "idiopathic" once the cause has been identified.

**Multiple Idiopathic Cervical Root Resorption (MICRR): A Brief Introduction**

MICRR affects multiple teeth within the dentition [8; 9; 10; 11]. MICRR lesions are often asymptomatic, non-carious, and lack overt gingival inflammation, increased pocket depth, or tooth mobility associated with classical cases of periodontal disease. Histologically, numerous resorptive areas are noted along root surfaces with evidence of osteoclasts/odontoclasts contained in Howship's lacunae [11; 12; 13]. MICRR lesions are frequently aggressive in nature and resistant to interventions, ultimately resulting in tooth loss [11; 12; 13]. Often, MICRR is detected as an incidental finding on radiographs or during routine dental examination. Fortunately, with enhanced tools and technologies over the last decade [14; 15], our understanding of MICRR etiology and its course has been improving, which will result in refining clinical management and thus, better outcomes.

While idiopathic root resorption is considered a rare condition, it is one that many dentists encounter over years of practice. The effects are devasting, e.g., loss of dentition, a feeling of helplessness by clinicians and patients due to lack of effective prevention/treatment options, and poor esthetics and function driving patients into isolation, negatively impacting mental health and wellbeing. Since the first case of MICRR described by Mueller and Rony in 1930 [16], etiologies of MICRR have been largely speculative. Here, we provide an overview of the cellular and molecular mechanisms mediating root resorption, and provide examples of etiology, including systemic conditions and genetic factors, medications, viral infections, inflammatory conditions, environmental and other proposed causes for MICRR [17].

**Cellular and Molecular Mechanisms of Root Resorption**

Histological studies in humans and animals have unequivocally demonstrated that root resorption is mediated by osteoclasts/odontoclasts and is distinct from bacterial-mediated cariogenesis [5; 7; 11; 12; 13; 18]. Yet the factors that activate osteoclasts/odontoclasts and recruit them to root surfaces rather than bone surfaces (noted in periodontitis) remain unknown. Bernhard Gottlieb in 1923 noted cases of periodontal disease that were not associated with marked inflammation but rather with perceived defective cementum formation [19; 20; 21]. He observed that "cementum was the only tissue which connects tooth with the body", and if not formed correctly, would put individuals at risk for a type of periodontal disease (characterized by gingival recession or pocket formation), which he termed as "marginal cementopathia" [19; 20]. Clinicians and researchers have revisited this concept, especially with increased knowledge about conditions that trigger clastic activity, including those associated with defective cementum formation [21; 22].

During initiation of cervical root resorption, the portal of entry is the cementum below the gingival epithelium, and resorption starts with localized destruction and/or removal of PDL [18]. Response to PDL injury includes formation of a blood clot and inflammation, followed by granulation tissue and recruitment of macrophages to the affected area [18; 23]. Impaired vasculature in the area leads to hypoxia, which promotes osteoclast differentiation and activity [24]. As the osteoclastic/odontoclastic resorptive lesions expand towards the pulp space by destroying cementum, dentin, and enamel, several resorption channels and interconnections with PDL (portal of exits) are created, generating a 3D space [18; 25]. In most cases, the advancing resorptive lesions are prevented from perforating into the pulp space by the pericanalar resorption resistant (PRSS) sheet [18; 25]. This layer consists of predentin, dentin, and occasionally reparative bone-like tissue. In the final stages of the disease, repair and remodeling sometimes occurs through the activity of cementoblast/osteoblast-like cells, resulting in deposition of bone-like tissue into the resorption cavities [18; 25].

## SYSTEMIC CONDITIONS AND GENETIC FACTORS

## Systemic Conditions and Syndromes

Root resorption has been associated with systemic and syndromic conditions, including endocrine disorders. In some of these conditions, dysregulated resorption affects the skeleton, leading to reduced bone mineral density or osteolytic lesions. Examples of conditions associated with root resorption include: hypothyroidism [26], hyperparathyroidism [27; 28], systemic sclerosis [29], Gaucher's disease [30], hereditary hemorrhagic telangiectasia [31], Paget's disease of bone [32; 33], Goltz syndrome [34], Papillon–Lefévre syndrome [35], and Turner syndrome [36]. To date, no studies have firmly established causality between these conditions and MICRR.

## Altered Osteoclast/odontoclast Activity

Familial expansile osteolysis (FEO; OMIM#174810) is caused by mutations in the *TNFRSF11A* gene [37; 38; 39], which encodes for receptor activator of nuclear factor κ-B (RANK), a receptor found on osteoclasts and its progenitors. Upon binding to RANK ligand (RANKL), RANK promotes osteoclastic formation/function. In FEO, *TNFRSF11A* mutations affecting its signaling peptide may result in constitutive activation independent of RANK ligand stimulation and uncontrolled osteoclast activity [38]. Individuals with FEO often present with early-onset deafness, skeletal deformities, and premature loss of teeth [38; 39; 40]. FEO has been associated with extensive resorption of cervical and apical areas of permanent teeth [41; 42]. Recently, Hajishengallis and colleagues reported a case of FEO in a 10-year-old female with missing ossicles and MICRR [37]. In addition to MICRR affecting at least 7 erupted permanent teeth, premature atypical root resorption of all primary teeth (started at age 5 and progressed with most of the roots resorbed by age 7) and resorption of an unerupted permanent canine was noted (**Fig**

**1A-D**). Genetic testing focusing on missing ossicles at the time of birth was inconclusive and the accelerated root resorption of primary teeth was not well appreciated. However, when the aggressive root resorption involved permanent teeth, it prompted further endocrinology and genetic testing, which revealed decreased lumbar spine mineral density, high circulating alkaline phosphatase [43] levels, and identification of the *TNFRSF11A* mutation, which together led to the FEO diagnosis.

MICRR in this case involved several anterior teeth (maxillary lateral incisors, left central incisor, unerupted left canine, and mandibular central incisors and left canine). Interestingly, cone-beam computed tomography scans of the resorptive defects suggested that the lesions started from a small portal of entry in the cementum and expanded below the bone level inside the tooth sparing the pulp canal space. Unfortunately, two of the affected teeth (mandibular lower central incisors) showed increased sensitivity and required extraction. MicroCT analysis of these teeth revealed defective formation of root cementum (**Fig 1E)**. Other potential complications of FEO include progressive osteoclastic resorption that can lead to severe, painful, disabling deformities, and pathologic fractures of bones. Although the patient in this case showed only a possible mineralization disorder, she was placed on bisphosphonates for management of overall skeletal problems. Ten months later, her biochemical markers of the disease were reversed, and the root resorption lesions appeared stable. The transdisciplinary medical-dental collaboration between Children's Hospital of Philadelphia and University of Pennsylvania School of Dental Medicine, generated sufficient diagnostic information to identify the cause of this young patient's condition, leading to appropriate and effective medical and dental treatment.

Genetic variants linked specifically with root resorption, but not overt systemic/syndromic manifestations have been rare to date. A few reports suggest a genetic predisposition to MICRR based on hereditary patterns [12; 15; 44], and several reports noted MICRR in healthy individuals with apparently non-contributory medical histories [8; 9; 10; 11; 13]. Neely, Thumbigere-Math, and colleagues reported a familial pattern of MICRR with a 30-year follow-up [12; 13]. To the best of our knowledge, this is the only report of inherited MICRR with an extended follow-up. The family included two generations with four MICRR-affected and four unaffected family members [12; 13]. The 63-year-old proband presented with a history of MICRR affecting multiple teeth (**Fig 1F-H**). Over several decades, the resorptive lesions progressed with a total of 19 affected teeth, leading to extraction/exfoliation of 12 teeth. Additionally, the proband's two sons and one daughter developed MICRR during their fourth to sixth decades of life. All affected subjects were asymptomatic, lacked known predisposing factors, and reported a non-contributory medical history. Whole exome-sequencing identified a novel autosomal dominant heterozygous mutation (c.1219 G>A (G388S) in the interferon regulatory factor 8 (*IRF8*) gene, which encodes a transcription factor that negatively regulates osteoclast differentiation [15]. *In vitro* and *in vivo* functional analysis demonstrated that IRF8$^{G388S}$ mutation promoted increased osteoclastogenesis, thus providing a molecular basis for enhanced root resorption. Based on MICRR-associated variants in *TNFRSF11A* and *IRF8*, other variants targeting key regulatory steps in the osteoclast/odontoclast pathway might increase predisposition to root resorption. This concept has been borne out by studies using a transgenic mouse model where knockout of *Tnfrfsf11b* (osteoprotegerin), a decoy receptor for RANKL, promoted extensive molar root resorption [45].

**Altered Periodontium as a Contributing Factor**

Defective cementum formation has been suggested to predispose to periodontal breakdown, i.e., the concept of "periodontosis" by Gottlieb [21]. Reduction or absence of acellular cementum at the cervical root surface theoretically exposes the root to resorption. Mutations in tissue-nonspecific alkaline phosphatase (TNAP), encoded by the *ALPL* gene, result in the inherited mineralization defect, hypophosphatasia (HPP; OMIM#146300, 241500, 241510) [46; 47; 48]. Early exfoliation of deciduous teeth and loss of permanent teeth are pathognomonic signs of HPP due to defective cementogenesis. Abnormal root resorption in permanent teeth of some HPP patients has been reported [49; 50], possibly associated with cementum defects. Other inherited cementum defects in humans are rare, but genetically engineered mouse models serve as proof-of-principle examples. Mice deficient in bone sialoprotein (BSP), an extracellular matrix protein critical for cementum mineralization and function, exhibit a lack of acellular cementum and subsequent periodontal breakdown [51; 52]. BSP null mice feature dramatic osteoclast/odontoclast mediated root resorption exclusively targeting the cervical regions of all molars (**Fig 1 I, J**). The cementum defect and periodontal destruction in the absence of inflammation illustrate Gottlieb's periodontosis concept and suggest other inherited periodontal structural defects may promote cervical root resorption.

Inherited defects likely intersect with acquired or environmental factors to increase susceptibility to MICRR, possibly explaining delayed onset and diagnosis of root resorption in some cases. We emphasize that not all cases are associated with genetic etiologies, although increased understanding of genetic inputs in MICRR should prompt genetic testing when cases cannot be explained by local etiologic factors or systemic abnormalities.

# MEDICATION-INDUCED ROOT RESORPTION

It is well recognized that medications have side-effects that can adversely affect oral health. For example, certain medications used to treat epilepsy, hypertension, and heart disease, or immunosuppressants in organ transplant patients, are associated with gingival hyperplasia [53; 54]. Several medications cause severe xerostomia (dry mouth) [55; 56]. Yet other therapies cause severe oral mucositis requiring treatment alterations. The examples below underscore the importance of collecting detailed medication histories (e.g., prescribed treatments as well as mouth rinses, toothpastes, herbal products, and vitamins) in individuals exhibiting root resorption. Significantly, these situations serve to remind physicians to consider treatment effects on oral health and incorporate dental clinicians in monitoring overall health of patients.

## Anti-resorptive Medications and Potential Rebound Effect Associated with MICRR

Anti-resorptive therapies are widely prescribed for treatment of osteoporosis and painful osteolytic manifestations of cancer. Several generations of bisphosphonates have served as key anti-resorptive agents for decades, while more recent therapies target regulators of osteoclast, osteocyte and osteoblast differentiation and function. Denosumab is a monoclonal anti-RANKL antibody that inhibits RANK-mediated activation of osteoclasts. Recently, Deeb and colleagues reported that a 69-year-old patient who discontinued denosumab after 5 years experienced MICRR affecting multiple teeth (**Fig 1K**), in conjunction with pain and sensitivity, but no alterations in attachment levels  [57]. A surge in osteoclastic activity may provide an explanation for occurrence and progression of MICRR, i.e., a rebound effect after discontinuing anti-resorptive therapy. After administration of denosumab, osteoclast activity rapidly declines and can drop by over 80% within weeks to months and remain at that level while denosumab treatment is continued [58]. Once treatment is discontinued, antibody levels suddenly decline, resulting in transient increases in osteoclastic activity and bone turnover to levels above the starting range, before eventually returning to pretreatment levels [59]. Although at this time there is insufficient evidence to support a causality between denosumab and MICRR, a website established to self-report cases of root resorption in patients treated with denosumab revealed 20 cases between 2013 and 2019, affecting predominantly females over 60 years taking denosumab for 2-5 years (https://www.ehealthme.com/ds/prolia/tooth-resorption/). Due to their antiresorptive effects on alveolar bone, bisphosphonates have been explored as potentially useful in preventing root resorption of replanted teeth [60] [61]. While use of bisphosphonates appears to be a potentially effective strategy in these contexts, caution should be taken due to its recent association with MICRR and more established links to medication-related osteonecrosis of the jaw (MRONJ) [62; 63; 64; 65].

**Chemotherapy**

Llavayol et. al reported that a 16-year-old female who received chemotherapy for ovarian cancer developed MICRR in 12 teeth, 9 years later [66]. While the authors discarded other possible etiologies and hypothesized a correlation between chemotherapy and defective cementum and PDL, they were not able to establish the etiology. An alternative interpretation is that the medications affected mineral homeostasis, potentially activating osteoclast activity along with destruction of cementum and PDL.

# PETS AND VIRAL INFECTIONS

Cats exhibit a high incidence of external root resorption, termed feline odontoclastic resorptive lesions (FORL), a disorder that strongly resembles MICRR in humans. The prevalence of FORL is around 29% to 60% [67; 68], and more commonly seen in domestic vs. wild cats, and often in females [69; 70]. The etiology of FORL remains unknown, however, the mechanisms and progression of osteoclastic/odontoclastic root resorption appears similar to humans (**Fig 1L**) [71]. Proposed risk factors include increasing age, diet low in magnesium/calcium and higher in vitamin D, and low frequency of teeth cleaning [69]. FORL has been associated with Feline herpes virus 1 (FeHV-1), and it has been speculated that transmission of FeHV-1 to humans can initiate MICRR [5; 6; 72; 73; 74; 75; 76]. Von Arx et al. reported four individuals with MICRR who had extended contact with cats and presented positive titers of neutralizing antibodies against FeHV-1 [72]. Similarly, Wu and colleagues reported a case of MICRR in an individual who had contact with cats [9].

Other reports have described patients with herpes zoster/shingles who developed MICRR in corresponding areas of nerve innervation. Solomon et al. reported cervical root resorption in two teeth in a 31-year-old female with a positive history of herpes zoster infection in the corresponding division of the maxillary trigeminal nerve [77]. Similarly, Ramchandani and Mellor reported MICRR in a 72-year-old female who presented with a 17 year earlier history of herpes zoster infection of the maxillary division of the trigeminal nerve [78].

## AREAS OF NOTE

Many other examples based on case reports and anecdotal experiences from clinicians and patients are worth mentioning since they may trigger additional thoughts regarding mechanisms mediating purported root resorption. While most examples of MICRR are non-inflammatory, we would be amiss if we did not mention inflammatory conditions associated with root resorption, such as severe periodontal disease, where marked inflammation triggers factors causing both osteoclast-mediated bone and cementum resorption. Additionally, osteoclast inhibitors such as denosumab and bisphosphonates have been associated with an acute-phase response and the release of proinflammatory cytokines, possibly explaining another underlying mechanism between antiresorptive medications and MICRR [79; 80].

Environmental factors are another area warranting consideration. Examples here relate to Gottlieb's findings discussed above that some types of periodontal disease were associated with perceived defective cementum formation rather than marked inflammation [19; 20; 21]. Individuals with minor defects in cementogenesis, whether related to genetic factors or exposure to environment toxins during tooth development, when exposed to periodontal pathogens or other local factors may be more susceptible to MICRR. Answers to this proposed mechanism of MICRR will require coordination among transdisciplinary researchers and clinicians as well as patients. Other environmental factors include marked trauma to the oral region, which is known to mediate osteoclast/odontoclast root resorption, usually localized to the affected area. Other examples of environmental factors, but regionally specific vs MICRR, include playing of wind instruments, parafunctional habits and previous orthodontic treatment, the latter usually limited to apical root resorption.

## CONCLUSION

Several valuable points are to be gained from this perspective:

1. The need for transdisciplinary approaches to improve health outcomes.

   While this perspective focuses on MICRR, it emphasizes the need for transdisciplinary approaches in order to improve diagnosis and subsequently treatment of diseases/conditions, including the potential to identify medications that may affect oral health. The specific examples of MICRR provided here demonstrate that oral conditions need to be considered in the context of the whole body in order to move away from silos, which is unfortunately evident within the disciplines of dentistry and medicine. We should move towards integrated systems approaches for research and treatment.

2. MICRR is associated with multiple conditions.

   The complex etiology of MICRR presented here provides important information for research and in clinical practice related to development, maintenance and regeneration of mineralized tissues.

3. MICRR remains a puzzle to solve.

   Further transdisciplinary research from basic science to clinical studies are needed to define the etiology of MICRR, understand mechanisms honing osteoclasts to tooth root surfaces vs surrounding bone, and develop treatments to arrest osteoclast activity and help repair tooth root structures versus extraction.

## CONFLICT OF INTEREST

The authors declare that the research was conducted in the absence of any commercial or financial relationships that could be construed as a potential conflict of interest.

## AUTHOR CONTRIBUTIONS

All authors have contributed significantly to the outline of topics, and in writing the perspective and approving the final version of the manuscript. Authors are listed in alphabetical order.

## FUNDING

This work was funded by K99AR073926 (to EYC), R01DE027639, R03DE028632, R03DE028411 (to BLF), NIAMS intramural research program funding (to MJS), and R00DE028439, R03DE029258, and start-up funds from the University of Maryland School of Dentistry (to VTM).

## ACKNOWLEDGMENTS

We thank the subjects and families for their participation in the studies described.  We thank Mr. Michael Chavez for micro-CT imaging in Figure 1G and Dr. Anthony Neely for the clinical photos in Figure 1 and his insights into the potential causes of MICRR based on his clinical experiences.

**Table 1. Potential Risk Factors for multiple idiopathic cervical root resorption (MICRR).**

| Potential Predisposing Factors | | Examples and References |
|---|---|---|
| Systemic Conditions and Genetic Factors | Systemic Diseases | Hypothyroidism [26], hyperparathyroidism [27; 28], Systemic sclerosis [29], Gaucher's disease [30], Hereditary hemorrhagic telangiectasia [31], Paget's disease of bone [32; 33] |
| | Syndromes | Goltz syndrome [34], Papillon–Lefe`vre syndrome [35], and Turner syndrome [36] |
| | Genetic Mutations & Altered Osteoclast/odontoclast Activity | *IRF8 [15]*, *TNFRSF11A* [37; 41; 42], *NFATC1*(unpublished data from our group) |
| | Genetic Mutations & Altered Periodontium | *ALPL* [49; 50], *BSP* [51; 52] |
| Medication-Induced Root Resorption | Antiresorptives | Discontinuation of Denosumab therapy [57], https://www.ehealthme.com/ds/prolia/tooth-resorption/ <br><br> Administration of bisphosphonate therapy may prevent MICRR progression [60] [61] |
| | Chemotherapeutics | [66] |
| Pets and Viral Infections | FeHV-1 | [5; 6; 9; 72; 73; 74; 75; 76; 81] |
| | Herpes Zoster | [77; 78] |
| Trauma | Fracture, Luxation, Replantation, Transplantation | [82; 83; 84; 85; 86] |
| Environmental Factors | Parafunctional Habits | Bruxism, Tongue Thrusting [25; 82; 84] |
| | Previous Orthodontic Treatment | Fixed appliances, Esthetic Brackets, Application of force greater than 20-26g/cm$^2$ [82; 84; 87] |
| | Musical Instruments | Wind instruments [88] |

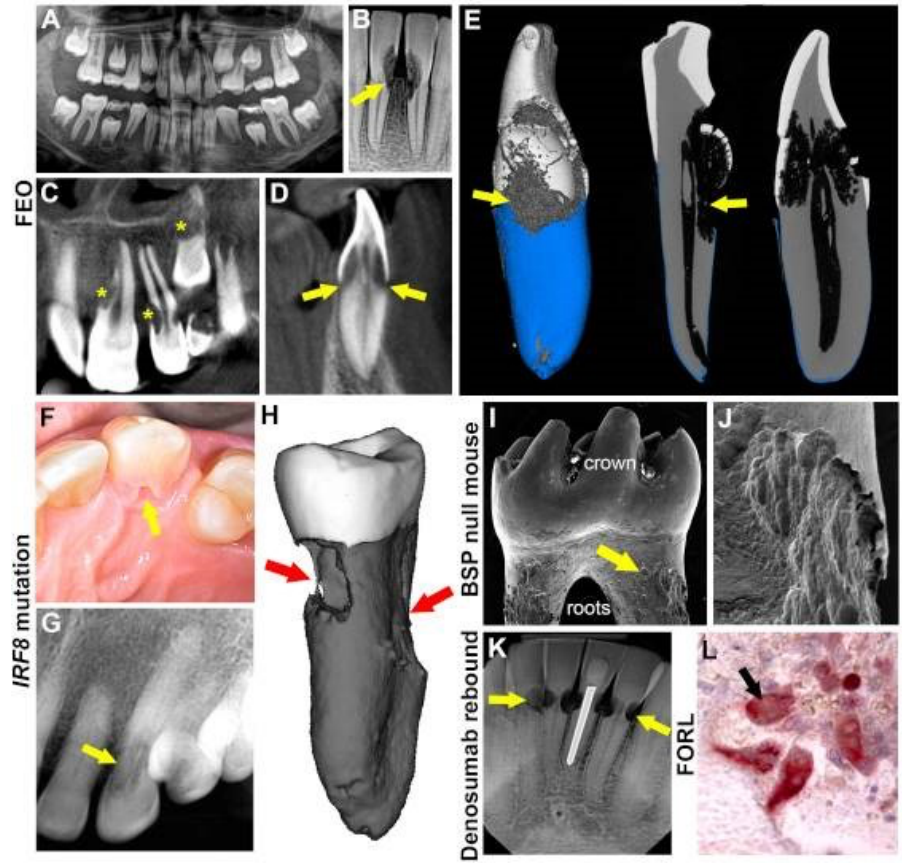

**Figure 1. Multiple etiologies of MICRR. (A)** Panoramic radiograph of 8-year-old female with familial expansive osteolysis (FEO) associated with *TNFRSF11A* variant showing cervical root resorption of permanent maxillary and mandibular incisors (yellow arrows). **(B)** Periapical radiograph of the same patient with FEO, at 9-years-old, showing extent of external resorption (yellow arrows). **(C, D)** Cone beam computed tomography (CBCT) of the same patient at 9-years-old showing extensive resorption of permanent central and lateral incisors (yellow stars) and unerupted canine (yellow star and yellow arrows). **(E)** 3D micro-computed tomography (micro-CT) reconstruction of incisor of the same FEO patient at age 9, noting defective cementum formation and root resorption. **(F)** Intraoral photograph of an advanced resorption lesion (yellow arrow) on the palatal aspect of tooth from affected individual with an inherited *IRF8* variant. **(G)** Radiograph of the lesion (yellow arrow) shown in panel F. **(H)** Micro-CT

reconstruction of an extracted tooth exhibiting extensive cervical resorption (red arrows). **(I)** Scanning electron microscopy [89] image of BSP null mouse molar showing abundant pitting (yellow arrows) at cervical root surfaces. **(J)** Higher magnification SEM of the tooth in panel I showing details of cervical root resorption in BSP null mouse molar. **(K)** Periapical radiograph of lower anterior teeth in 69-year-old patient, after discontinuation of denosumab, showing multiple areas of cervical root resorption (yellow areas). **(L)** Tartrate-resistant acid phosphatase (TRAP) stain of histology section showing multinucleated odontoclasts (red cells, black arrow) on root surfaces of a cat with feline odontoclastic resorption lesions (FORL). Panels A-D reproduced with permission from Macaraeg et al., *Pediatr Dent* 42(1):62-65, 2020. Panels F and G reproduced with permission from Neely et al., *J Periodontol* 87(4):426-433, 2016. Panel I reproduced with permission from Foster et al., *J Dent Res* 92(2):166-172, 2013. Panel K reproduced with permission from Deeb et al., *J Endod* 45(5): 640-644, 2019.